\documentclass[reprint,nofootinbib,amsmath,amssymb,prl]{revtex4-2}

\usepackage{graphicx}
\usepackage{dcolumn}
\usepackage{bm}
\usepackage{xcolor}

\begin{document}

\title{Spectral densities from Euclidean lattice correlators via the Mellin transform}

\author{Mattia Bruno}
\author{Leonardo Giusti}
\author{Matteo Saccardi}
\affiliation{Dipartimento di Fisica, Universit\`a degli Studi di  Milano–Bicocca and INFN, Sezione di Milano–Bicocca, Piazza della Scienza 3, I-20126 Milano,Italy}
                
\date{\today}

\begin{abstract}
Spectral densities connect correlation functions computed in quantum field theory to observables measured in experiments. For strongly-interacting theories, their non-perturbative determinations from lattice simulations are therefore of primary importance. They entail the inverse Laplace transform of correlation functions calculated in Euclidean time. By making use of the Mellin transform, we derive explicit analytic formulae to define spectral densities from the time dependence of correlation functions, both in the continuum and on the lattice. The generalization to smeared spectral densities turns out to be straightforward. The formulae obtained here within the context of lattice field theory can be easily applied or extended to other areas of research.

\end{abstract}

\newcommand{\eq}[1]{Eq.~(\ref{#1})}
\newcommand{\Eq}[1]{Eq.~(\ref{#1})}

\newcommand{\refs}{Hansen_2017,
Hashimoto_2017,
Bulava:2019kbi,
HLT,
Bulava:2021fre,
DelDebbio:2022qgu,
Boito:2022njs,
alexandrou_2023,
Bennett:2024cqv}
\newcommand{\refsmath}{laplace,pike,epstein}

\newcommand{\transitions}[0]{Lin:2001ek,Detmold:2004qn,Kim:2005gf,Christ:2005gi,Hansen:2012tf,Bernard:2012bi,Briceno:2014uqa,Briceno:2015csa,Briceno:2015tza,Baroni:2018iau,Briceno:2019nns}

\maketitle

\def\Id{\mathbb{I}}
\def\H{\mathcal{H}}
\def\Lrho{L^2(0,\infty,d\omega)}
\def\vbar{\overline{v}}

\textbf{\textit{Introduction}} --- The lattice regularization combined with Monte Carlo simulations allow to determine non-perturbative dynamical properties of relativistic field theories from first principles. 
The primary quantities estimated by numerical simulations (hence with statistical errors) are correlation functions of (composite) fields in Euclidean space-time calculated from path-integral averages. 
The physical information is contained in the so-called spectral densities $\rho(\omega)$, functions of the energy $\omega$ with support over $[\omega_0\! > \! 0,\infty)$ in theories with a mass gap.
Thanks to the K\"all\'en–Lehmann spectral decomposition, generic two-point correlators\footnote{We take the time-momentum representation of correlators, omitting the momentum dependence from the notation, and consider the zero momentum case, but our results are fully generic. Note that our derivation applies straightforwardly to higher $n$-point functions, involving multiple Euclidean-time separations.} $C(t)$ are related to the corresponding spectral densities by a Laplace transform 
\vspace*{-0.25cm}
\begin{equation}
    C(t) = \int_\omega \rho(\omega) \, e^{-\omega t} \,, \quad \text{with } \int_\omega \equiv \int_0^\infty d\omega \,.
    \label{e:ct}
\end{equation}
Since the direct analytic continuation back to Minkowski time is not a viable option, low-energy properties of $\rho(\omega)$ (including stable particles) or amplitudes at threshold~\cite{Maiani:1990ca,Bruno:2020kyl} are usually extracted from the large Euclidean-time behavior of correlators.
Lattice simulations, however, are performed on a finite volume where spectral densities become weighted sums of $\delta$ functions, whose weights have power-like finite-size effects when at least two particles are involved. 
The relation with their infinite-volume counterpart is known, up to exponentially small systematic errors, only in limited kinematic regions and for a small number of  particles~\cite{Luscher:1986pf,Meyer:2011um,Lellouch:2000pv,\transitions}; see Refs.~\cite{Briceno:2017max,Hansen:2019nir} for recent reviews and Refs.~\cite{RBC:2020kdj,Andersen:2018mau,Hansen:2020otl,Feng:2014gba} for examples of numerical calculations.

To overcome such limitations, thanks also to algorithmic and technological developments which unlocked unprecedented levels of precision in modern calculations, the direct extraction of smeared spectral densities from lattice data has been recently reconsidered and has received significant attention~\cite{Hansen_2017,HLT,Bailas:2020qmv}.
It entails the Inverse Laplace Transform (ILT) of correlators, a notorious mathematically ill-posed inverse problem, which becomes particularly difficult when dealing with discrete noisy data.
The practical solution adopted in state-of-the-art studies relies on the numerical Backus-Gilbert procedure and its generalizations~\cite{Hansen_2017,Bulava:2019kbi,HLT,Bailas:2020qmv,Gambino:2020crt,Bulava:2021fre,DelDebbio:2022qgu,Boito:2022njs,Gambino:2022dvu,alexandrou_2023,Barone:2023tbl,Bennett:2024cqv}.

Building on the vast mathematical literature on the Inverse Laplace Transform~\cite{\refsmath}, the aim of this letter is to derive explicit analytic formulae for defining spectral densities from the Euclidean-time dependence of the corresponding correlation functions, both in the continuum theory and at finite lattice spacing. 
The generalization to smeared spectral densities, defined from their convolution with known kernel functions $\kappa(\omega)$
\vspace*{-0.125cm}
\begin{equation}
    \rho_\kappa \equiv \int_\omega \rho(\omega) \, \kappa(\omega) \,,
    \label{eq:rhokappa}
\end{equation}
is also derived.
The explicit formulae presented here are not only relevant for practical computations, but they additionally pave the way to study the continuum and infinite-volume limits by using Symanzik's~\cite{Symanzik:1983dc,Symanzik:1983gh,Symanzik:1979ph} and L\"uscher's~\cite{Luscher:1985dn,Hansen:2019rbh,Hansen:2020whp} analyses respectively. In particular, the presence of a smooth smearing kernel (e.g. Gaussian or Cauchy types) may be beneficial when the difference between the finite- and the infinite-volume $\rho_\kappa$ vanishes fast enough (exponentially)~\cite{BHprep}.

Once this program will have been completed, direct non-perturbative predictions of spectral densities will be relevant for a better comprehension of the theory itself, e.g. its particle and resonance content.
Smeared spectral densities from physics-motivated kernels will give access to phenomenologically relevant observables, such as inclusive hadronic cross sections, semi-leptonic decay rates, or non-static properties of the quark-gluon plasma, to name a few (see for instance Refs.~\cite{Hansen_2017,Gambino:2020crt,Jeon:1995zm,Meyer:2007ic}).

The results derived here can be directly applied or generalized to other fields of research and applied science, when the desired target observable is the ILT of what is effectively measured.

\vskip 2ex
\textbf{\textit{Preliminaries on the Laplace transform}} --- Based on known results from the mathematical literature~\cite{\refsmath}, in this section we introduce the generic solution to the ILT in the continuum.
We consider a vector space of real functions with support 
over $\mathbb R^+$ equipped with the inner
product $\langle f \vert g \rangle = \int_0^\infty dt \, f(t) \, g(t)$.
Denoting by $\Id$ the identity operator, we introduce the orthonormal 
\emph{time basis} $\{ \vert t \rangle \}$, with $t \in [0,\infty)$, 
such that $\langle t \vert t' \rangle = \delta(t-t')$ and 
\vspace*{-0.25cm}
\begin{equation}
    \Id = \int_t \vert t \rangle \langle t \vert \,, \quad \text{with }
    \int_t \equiv \int_0^\infty dt \,.
    \label{eq:tbasis}
\end{equation}
Next, we consider a second set of (non-orthogonal) vectors $\{ \vert \omega \rangle \}$, with $\omega \in [0, \infty)$, whose coordinates in the time basis are
\vspace*{-0.25cm}
\begin{equation}
    \langle t \vert \omega \rangle = \langle \omega \vert t \rangle = e^{-t\omega} \,.
    \label{eq:omegat}
\end{equation}
Note that in our notation both $t$ and $\omega$ are dimensionless variables, and when they carry a physical dimension, e.g. time and energy, we assume to take a common reference scale to cancel it.
A consequence of \eq{eq:omegat} is that the metric 
\vspace*{-0.25cm}
\begin{equation}
    \langle \omega \vert \omega' \rangle = 
    \frac{1}{\omega+\omega'} \equiv \H(\omega,\omega')
    \label{eq:H}
\end{equation}
is non-diagonal and corresponds to the so-called Carleman operator~\cite{carleman}.
With the aim of diagonalizing $\H(\omega,\omega')$, we introduce the ``Mellin'' basis $\{ \vert s \rangle \}$ with $s \in \mathbb R$, defined as~\cite{laplace,pike,epstein}
\begin{equation}
    u_s(t) = \langle t \vert s \rangle \equiv \frac{e^{is \log(t)}}{\sqrt{2\pi t}}\, .
    \label{eq:us}  
\end{equation}
As shown in Ref.~\cite{inprep}, the vectors $\{ \vert s \rangle \}$ form an orthonormal basis, and obey the relation
\begin{equation}
 \langle \omega \vert s \rangle  =  \int_t \, e^{-tw} \, u_s(t) = \lambda_s \, u_{-s}(w) = \lambda_s \, u_s^\ast(w) \,,
    \label{eq:LT_us}
\end{equation}
where $\lambda_s \equiv \Gamma\left(\frac12+is \right) $ and $\Gamma(z)$ is the Euler Gamma function. 
In this basis the metric $\H(\omega,\omega')$ is diagonal
\begin{equation}
\H(\omega,\omega')
    = \int_s \langle \omega \vert s \rangle \langle s \vert \omega' \rangle = \int_s u_s^\ast(\omega) \,  \vert \lambda_s \vert^2 \, u_s(\omega') \,,
    \label{eq:Hdiagonal}
\end{equation}
with eigenvalues 
$\vert \lambda_s \vert^2 = \frac{\pi}{\cosh(\pi s)} \in \left[ 0, \pi \right]$ and $\int_s = \int_{-\infty}^\infty ds$.
From \eq{eq:LT_us} it follows that the Laplace transform is bounded and self-adjoint with purely absolutely 
continuous spectrum, $\pm \vert \lambda_s \vert \in [ -\sqrt\pi,+\sqrt\pi ]$, of multiplicity one.
Similarly, we observe that $\H(\omega,\omega')$ is bounded, with spectrum of multiplicity two, acting on functions in $\Lrho$.

Analogously to the Fourier-transform, despite the representation in terms of the complex functions $u_s(\omega)$, the integral in \eq{eq:Hdiagonal} is real.
In fact, an equivalent diagonal representation with real basis functions exists~\cite{inprep}.

\begin{figure}
\centering
\includegraphics[width=.45\textwidth]{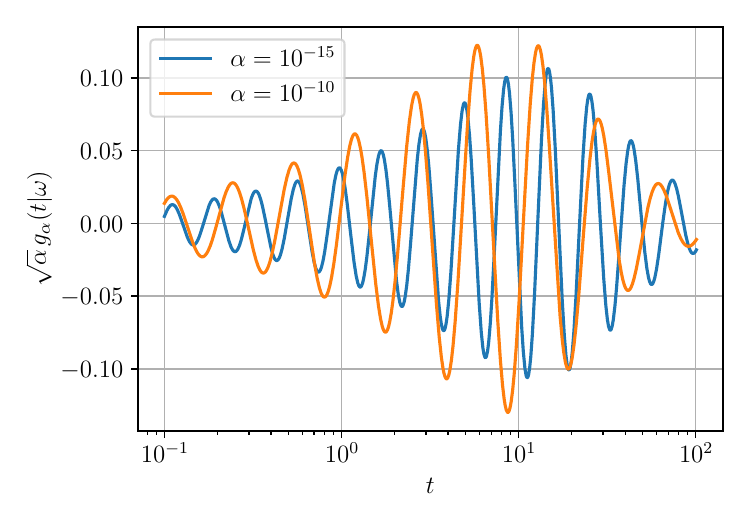}
\vspace{-0.5cm}

\caption{Coefficients as in \eq{eq:gt} rescaled with $\sqrt\alpha$ at fixed $\omega=0.5$, as a function of time. Notice the logarithmic scale along the horizontal axis.
\label{fig:coeffs}}
\end{figure}

\vskip 2ex
\textit{Inverse Laplace transform} --- Let us consider functions $C(t)$ in $L^2(0,\infty,dt)$, satisfying the representation in \eq{e:ct}, namely
\vspace*{-0.25cm}
\begin{equation}
    \vert C \rangle = \int_\omega \rho(\omega)\, \vert \omega \rangle\,. 
    \label{e:C}
\end{equation}
When only $\vert C \rangle$ is known, the spectral density is obtained from the solution of the Fredholm integral equation of first kind
\vspace*{-0.125cm}
\begin{equation}\label{eq:omegaC}
    \langle \omega \vert C \rangle = \int_{\omega'} \H(\omega,\omega') \, \rho(\omega') \,.
\end{equation} 
According to Hadamard, the problem is mathematically ill-posed (for additional details see for example Ref.~\cite{groetsch1984theory}), due to the difficulties in inverting $\H(\omega,\omega')$ stemming from the rapid exponential decay of its eigenvalues, which tend to zero in the limit $s \to \pm \infty$.
This does not imply the absence of a solution, since its inverse may nevertheless be defined by regulating the smallest $\vert \lambda_s \vert^2$ with the introduction of the Tikhonov-regulated operator~\cite{tikhonov,tikhonov1,groetsch1984theory}
\begin{equation}
    \H^{-1}_\alpha(\omega,\omega') = \int_s u_s^\ast(\omega) \frac{1}{\vert \lambda_s \vert^2 + \alpha} u_s(\omega') \,, \quad \alpha>0 \,.
    \label{eq:Hinv}
\end{equation}
Its convolution with $\H(\omega,\omega')$ forms the smeared Dirac $\delta$ function
\begin{equation}
    \delta_\alpha(\omega,\omega') = \langle \omega \vert S_\alpha \vert \omega' \rangle = \int_s u_s^\ast(\omega) \frac{\vert \lambda_s\vert^2}{\vert \lambda_s \vert^2 + \alpha} u_s(\omega') \,,
    \label{eq:delta_alpha}
\end{equation}
obeying $\lim_{\alpha \to 0} \delta_\alpha(\omega,\omega') = \delta(\omega - \omega')$ in the space $\Lrho$~\cite{inprep}, with the operator $S_\alpha$ defined as
\begin{equation}
    S_\alpha \equiv \int_s \vert s \rangle \frac{1}{\vert \lambda_s \vert^2 + \alpha} \langle s \vert \,.
    \label{eq:Salpha}
\end{equation}
Defining the convolution of the spectral density with $\delta_\alpha(\omega,\omega')$ as $\rho_\alpha(\omega)$, the ILT is obtained from the limit
\begin{equation}
    \rho(\omega) = \lim_{\alpha \to 0} \rho_\alpha(\omega) \,, 
    \label{e:rho_sol}    
\end{equation}
with $\rho_\alpha(\omega)$ being
\begin{equation}
    \rho_\alpha(\omega)= \langle \omega \vert S_\alpha \vert C \rangle = \int_t g_\alpha(t|\omega) \, C(t) \,.
    \label{eq:rhosol}
\end{equation}
Since in practice we know $\vert C \rangle$ only in the time basis, in the equation above we inserted the completeness relation arriving at the coefficients
\begin{equation}
    g_\alpha(t|\omega) = \langle \omega \vert S_\alpha \vert t \rangle = \int_s \frac{u_s^\ast(\omega) \lambda_s u_s^\ast(t)}{\vert \lambda_s \vert^2 + \alpha} \,. 
    \label{eq:gt}
\end{equation} 
The $g_\alpha(t|\omega)$ are real ($s \in \mathbb R$) and, for finite $\alpha$, they can be calculated thanks to the rapid fall-off of the numerator at large $s$~\cite{murli}, see Fig.~(\ref{fig:coeffs}) for a concrete example.

By expressing the smeared $\delta$ function in terms of the coefficients in \eq{eq:gt}, $\delta_\alpha(\omega,\omega') = \int_t \langle \omega \vert S_\alpha \vert t \rangle \langle t \vert \omega \rangle$, one may show that they minimize the regularized norm $\int_\omega [\delta(\omega-\omega') - \delta_\alpha(\omega,\omega')]^2 + \alpha \langle \omega' \vert S_\alpha^2 \vert \omega' \rangle$~\cite{inprep,groetsch1984theory,power1980}.
Similarly our definition in \eq{eq:rhosol} minimizes the norm $\int_\omega [\rho(\omega) - \rho_\alpha(\omega)]^2 + \alpha \langle C \vert S_\alpha^2 \vert C \rangle$, see Ref.~\cite{inprep}.

\vskip 2ex
\textbf{\textit{Spectral densities from Euclidean correlation functions}} --- In relativistic field theories, the convergence and uniqueness of the solution above is guaranteed by our a priori theoretical knowledge\footnote{In the absence of such prior knowledge, one should refer to the most general conditions under which $C(t)$ may be interpreted as the Laplace transform of another function~\cite{shilov}.} on the validity of the representation\footnote{Depending on the quantum numbers, when several densities appear in the parametrization of the correlator, one should consider the appropriate linear combinations to satisfy \eq{e:ct}.} in \eq{e:ct}.
For theories with a mass gap, the infrared regime of spectral densities is automatically regulated. Instead care is required at small $t$ due to short-distance singularities, typical of integrated correlators, and modifications of \eq{eq:gt} are in general mandatory for quantum field theories, where $\rho(\omega) \in \Lrho$ is often not satisfied. 
For instance, for the isovector vector correlator in QCD at zero spatial momentum,
$C(t)$ in \eq{e:ct} scales proportionally to $1/t^3$ at short times.

\def\rhos{\rho_{\mathsf s}}
\vskip 2ex
\textit{Regulated spectral densities} --- Let us consider a correlator $C(t)$ obeying \eq{e:ct} and \eq{e:C}, with $\rho(\omega) = \omega^k \rhos(\omega)$ and $\rhos \in \Lrho$.
Inspired by dispersive relations, where the asymptotic growth is regulated by inverse powers of the energy with the addition of subtraction constants, we first generalize \eq{eq:LT_us} to 
\begin{equation}
    \int_t e^{-\omega t} \, t^p \, u_s(t) = \frac{\lambda_{s,p}}{\omega^p} u_s^\ast(\omega) \,,
\end{equation}
with $p \!>\! -\tfrac12 $ and the modified pre-factors $\lambda_{s,p} = \Gamma\left(\frac12 + p + i s \right)$.
Then we introduce the generalized coefficients
\vspace*{-0.25cm}
\begin{equation}
    g_\alpha(t|\omega,p,q) \equiv \int_s \frac{u_s^\ast(\omega) \, \lambda_{s,p} \, u_s^\ast(t)}{\lambda_{s,p} \lambda_{s,q}^\ast + \alpha} \, t^q\,,
    \label{eq:galpha_pq}
\end{equation}
for which \eq{eq:rhosol} turns into a valid representation of the spectral density $\rhos(\omega)$ for $q=k$ and $\forall p > -\frac12$. For the vector correlator in QCD mentioned before, $q$ should be larger than $5/2$ for the integral in \eq{eq:rhosol} to converge with $q=3$ being the most natural choice.
Considering spectral densities $\rhos(\omega)$ further suppressed by inverse powers of $\omega$ larger than $k$ may be beneficial in some practical calculations, and one may experiment with the values of $p$ and $q$ accordingly.

\vskip 2ex
\textit{Smeared spectral densities} --- As motivated earlier, in quantum field theories smeared spectral densities are of great interest\footnote{In general, $\kappa(\omega)$ (and $\rho_\kappa$ as a consequence) may depend on additional kinematic variables, like the mass of the decaying particle for rates, or the center of the Gaussian kernel for smeared densities. For better readability, we omit them from the notation.}.
For their extraction one treats the smearing kernel in complete analogy with \eq{eq:omegat}, i.e. by replacing $\vert \omega \rangle$ in \eq{eq:rhosol} with the vector
\begin{equation}
    \vert \kappa \rangle \equiv \int_\omega \kappa(\omega) \, \vert \omega \rangle \,.    
\end{equation}
The smeared spectral density is then simply found from the limit $\rho_\kappa = \lim_{\alpha \to 0} \rho_{\kappa,\alpha}$ with
\begin{equation}
    \rho_{\kappa,\alpha} = \langle \kappa \vert S_\alpha \vert C \rangle = \int_t g_{\kappa,\alpha}(t) \, C(t) \,,
    \label{eq:rhokappa_sol}
\end{equation}
and $g_{\kappa,\alpha}(t) = \int_\omega \kappa(\omega) \, g_\alpha(t|\omega)$. Clearly smeared regulated spectral densities may be obtained by trivially merging the two methods.

Before addressing the discrete case, we remark that \eq{eq:rhosol} and (\ref{eq:gt}) unlock the possibility to address directly in the continuum finite-size effects of (smeared) densities from those analytically known for the correlators.

\def\tmax{t_\mathrm{max}}

\vskip 2ex
\def\Hbar{\overline{\H}}

\textbf{\textit{Spectral densities from lattice correlation functions}} ---  In lattice field theories, $C(t)$ is sampled at discrete Euclidean times $t = an$, where $n \in \mathbb N$ and $a$ is the lattice spacing. Additionally the correlator at finite $a$, $\overline C_a(t)$, suffers from discretization errors, typically of $O(a^2)$, which we do not address here. When numerical simulations are used to estimate $\overline C_a(t)$, $n$ is limited to $n \in [0,\tmax/a]$, and $\overline C_a(t)$ is affected by statistical errors. In the following we discuss how to generalize to the lattice the results derived so far in the continuum.

\vskip 2ex
\textit{Discretization} --- 
By excluding the first time slice $t=0$ without loss of generality, the discretization procedure modifies the metric to
\vspace*{-0.125cm}
\begin{equation}
    \Hbar_a(\omega, \omega') = \frac{a e^{-a (\omega + \omega')}}{1 - e^{-a (\omega + \omega')}} \,,
    \label{eq:Ha}
\end{equation}
and accordingly also the inverse problem into 
\begin{equation}
    a\sum_{n=1}^\infty \langle \omega \vert an \rangle \langle an \vert \overline C_a \rangle = \int_{\omega'} \Hbar_a(\omega,\omega') \, \overline \rho_a(\omega') \,.
\end{equation}
where $\overline \rho_a(\omega)$ is the spectral density at finite $a$. Following the rationale of the approach in the continuum, we diagonalize (and invert) directly $\Hbar_a(\omega,\omega')$.
By starting from the analytic expression of the eigenfunctions of the (infinite) Hilbert matrix~\cite{Hill}, we find~\cite{inprep} that $\Hbar_a(\omega, \omega')$ shares the same eigenvalues $\vert \lambda_s \vert^2$ encountered above, with $s \in \mathbb R^+$. 
Its normalized eigenfunctions
\begin{equation}
\begin{split}
    v_s(\omega,a) \equiv & \, \sqrt{2\pi a} \frac{u_s(1-e^{-a\omega})}{\vert N_s \vert} e^{-a \omega} \vert \lambda_s \vert^2 \\& \times  {}_2F_1 \left( \begin{array}{c|}
      \frac12+is,\frac32+is    \\
       2
    \end{array} \, e^{-a\omega} \right) \,,
    \label{eq:vs}
\end{split}
\end{equation}
where ${}_qF_p$ are the hypergeometric functions~\cite{Gradshteyn:1702455} and
\begin{equation}
    N_s = \sqrt{2\pi} \frac{\Gamma(-2is) \lambda_s}{(\frac12-is) \lambda_s^\ast} \,,
    \label{eq:Ns}
\end{equation}
form an orthonormal basis in $\Lrho$, where $\Hbar_a(\omega,\omega')$ is diagonal, and such that a smeared $\delta$ function at finite $a$ may be introduced by replacing $u_s(\omega)$ with $v_s(\omega,a)$ in \eq{eq:delta_alpha}
\begin{equation}
    \overline \delta_{a,\alpha}(\omega,\omega') \equiv \int_{s\in \mathbb R^+} v_s(\omega,a) \frac{\vert \lambda_s \vert^2}{\vert \lambda_s \vert^2 + \alpha} v_s(\omega',a)\,.
\end{equation}
The spectral density at finite lattice spacing, $\overline \rho_a(\omega)$, is then recovered, on a lattice with infinite time extent, from the $\alpha \to 0$ limit of~\cite{inprep}
\begin{equation}
    \overline \rho_{a,\alpha}(\omega) \equiv a \sum_{t=a}^\infty \overline g_{a,\alpha}(t | \omega) \, \overline C_a(t) \,,
    \label{eq:rhosol_sum}
\end{equation}
with the coefficients
\begin{equation}
    \overline g_{a,\alpha}(t|\omega) = \int_{s\in \mathbb R^+} v_s(\omega,a) \frac{\vert \lambda_s \vert}{\vert \lambda_s \vert^2 + \alpha} \vbar_s(t-a,a) \,,
    \label{eq:gtalpha_bar}
\end{equation}
and~\cite{inprep}
\vspace*{-0.125cm}
\begin{equation}
    \vbar_s(t,a) \equiv  \frac{ \vert \lambda_s \vert^3}{\sqrt a \vert N_s \vert} \ {}_3F_2 \left( \begin{array}{c|}
      -\tfrac{t}{a},\tfrac12+is,\tfrac12-is    \\
       1,2 
    \end{array} \, 1 \right) \,.
\end{equation}
By expanding the eigenfunctions around $a=0$ we find
\begin{equation}
    v_s(\omega,a) = \sqrt a \,  \frac{2\mathrm{Re} [u_s(a\omega) N_s]}{\vert N_s \vert} \big(1 + O(a^2 \omega^2)\big) \,,
    \label{eq:cont_lim_vs}
\end{equation}
which also highlights the connection between the continuum basis and the discretized one.
Therefore, the solution in \eq{eq:rhosol_sum} is a discretized form of \eq{eq:rhosol}, but with new coefficients differing from $g_\alpha(t|\omega)$ by cutoff effects starting at $O(a^2)$. 
Despite being particularly pronounced at small times for the individual coefficients, the discretization effects induced by the ILT on the spectral density itself are instead minimized by $\overline g_{a,\alpha}(t|\omega)$, contrary to $g_\alpha(t|\omega)$. This is understood by noting that the coefficients in \eq{eq:gtalpha_bar} minimize the functional $\int_\omega [\overline \rho_a(\omega) - \overline \rho_{a,\alpha}(\omega)]^2$, up to regularization terms proportional to $\alpha$, in analogy to the continuum.
Indeed, in the $\alpha \to 0$ limit, $\overline \delta_{a,\alpha}(\omega,\omega') \to \delta(\omega-\omega')$ and $\overline C_a(t)$ is the only remaining source of discretization errors.
This is not the case if one takes \eq{eq:rhosol_sum} with $g_\alpha(t|\omega)$, since additional cutoff effects induced by the ILT survive as $\alpha \to 0$.
Intuitively, this may be expected from the wild short-distance oscillations of the continuum coefficients reported in Fig.~\ref{fig:coeffs}. 
This phenomenon is less extreme for other integral quantities, e.g. the Hadronic-Vacuum-Polarization contribution\footnote{Also for this quantity, it would be interesting to explore a similar philosophy to test a possible reduction of discretization errors.} to $(g-2)_\mu$~\cite{Bernecker_2011}.

Finally we note that when the Hilbert matrix is truncated at asymptotically large $\tmax/a$, the coefficients numerically calculated using the least-square approach, e.g. Ref.~\cite{HLT}, tend to our integral representation in \eq{eq:gtalpha_bar}. Extensions to schemes beyond Tikhonov's~\cite{hadron2023} and to regularized spectral densities may be easily incorporated into our formalism~\cite{inprep}.

\vskip 2ex
\textit{Finite temporal domain} --- Now we turn to the case where the correlator $C(t)$ is known on a finite temporal range $t \in [0,\tmax]$, e.g. the typical time subdomain of a simulated lattice where the non-zero temperature effects are negligible.
With this restriction the metric becomes
\begin{equation}
    \frac{1 - e^{-\tmax (\omega + \omega')}}{\omega + \omega'} \,.
    \label{eq:HT}
\end{equation}
Here the relevant observation is that, in field theories with a mass gap, the large-time behavior of correlation functions is bounded by a scale $M>0$ such that $0 \leq |C(t)| \leq |C(\tmax)| e^{-M (t-\tmax)}$ for $t>\tmax$.
As a consequence, the systematic difference between $\rho_\alpha(\omega)$ and the spectral density calculated by restricting the time interval in \eq{eq:rhosol} to $[0,\tmax]$ vanishes exponentially in $M \tmax$~\cite{inprep}.
Notice that this bound is not sufficient for the thermal theory. Following the strategy of this letter, one should instead diagonalize the appropriate finite-temperature metric, similar to the one in \eq{eq:HT}.

\vskip 2ex
\textit{Taming statistical errors} ---
In presence of statistical fluctuations, the ILT regulator may have a beneficial effect on the variance. Therefore it may be used to tame the statistical error on the spectral density, while keeping the corresponding systematic effect significantly below it. For instance, if one adopts Tikhonov's scheme, e.g. \eq{eq:Hinv}, by taking $\rho(\omega) \in \Lrho$ indeed one can show that
\begin{equation}
\begin{split}
    \vert \rho_\kappa-  \rho_{\kappa,\alpha} \vert^2 & \leq \, \alpha^2 \bigg[ \int_\omega \rho(\omega)^2 \bigg]  \\ & 
    \times \int_{\omega,\omega'} \kappa(\omega) \kappa(\omega') \int_s  \frac{u_s^\ast(\omega) u_s(\omega')}{(|\lambda_s |^2+\alpha)^2} \,.
    \label{eq:syst}
\end{split}
\end{equation}
This relation can be used to roughly estimate a good value of $\alpha$, but a test of the systematic effect due to $\alpha \neq 0$ has to be performed directly on the data, by varying the value of $\alpha$ in a suitable range.
This is not needed when the theoretical prediction is compared with the corresponding experimental data smeared with $\int_\omega \kappa(\omega) \delta_\alpha(\omega,\omega')$~\cite{Boito:2022njs,alexandrou_2023}.

\vskip 2ex
\textbf{\textit{Conclusions}} --- The level of precision reached by modern numerical lattice calculations may enable the direct computation of spectral densities. Motivated by this progress, we have derived the explicit analytic formula to relate them to Euclidean-time correlators in the continuum limit. 
An optimal adaptation to the discrete case would be achieved by replacing the integral over time with a sum spanning geometrically distributed (i.e. $a^\mathbb{N}$) data~\cite{Pike1984}, which however is evidently not usable in lattice field theories. 
To overcome this problem, we have generalized the formula for the continuum theory to the case of a regular discrete sampling in time, so that discretization errors induced by the ILT on the spectral density are minimized.

These findings open the door to computations with controlled and improvable systematic errors for a wide variety of phenomenologically relevant non-perturbative quantities. More specifically, finite-size and discretization effects on the estimated spectral density may be directly inferred from those of the correlators. 
In other fields where data is available under similar constraints, without the hassle of intrinsic discretization effects in $C(t)$, the solution in \eq{eq:rhosol_sum} is exact for $\alpha \to 0$.

\begin{acknowledgments}
We wish to thank N.~Tantalo for interesting discussions on the topic of this letter.
MS thanks M.T.~Hansen, L.~Del Debbio for stimulating conversations on the topic and the University of Edinburgh for hospitality during completion of this work. MB would like to thank M.T.~Hansen, C.~Lehner and T.~Izubuchi for previous and ongoing collaborations on related subjects. 
At the beginning of the project, MB was supported by the national program for young researchers ``Rita Levi Montalcini''.
This work is (partially) supported by ICSC - Centro Nazionale di Ricerca in High Performance Computing, Big Data and Quantum Computing, funded by European Union – NextGenerationEU.
\end{acknowledgments}

\appendix
\section{Supplemental Material}

To complement the discussion in the main text, we present here a few mathematical properties of basis vectors and operators relevant for the ILT and for regulated spectral densities. 

\vskip 2ex
\textbf{\textit{Properties of the ``Mellin'' basis}} ---  The orthogonality of the basis vectors $\{ \vert s \rangle \}$,  defined in \eq{eq:us}, may be proven by considering the change of variable $\eta=\log(t)$ so that
\begin{equation}
\hspace{-0.125cm}    \int_0^\infty dt \, u_s^\ast(t) \, u_{s'}(t)\! =\!
    \int_{-\infty}^{\infty} \frac{d\eta}{2\pi} \, e^{i \eta (s-s')}\! =\! \delta(s-s')\, ,
    \label{e:us_ortho}
\end{equation}
while the completeness of the basis is guaranteed by
\begin{equation}
    \int_s \, u_s^\ast(t) \, u_s(t') = \frac{\delta(\log(t/t'))}{\sqrt{tt'}} = \delta(t-t')\,.
    \label{e:us_comp}
\end{equation}
Analogously to the Fourier transform, also in this case it is possible to work with a basis of real eigenfunctions
\begin{align}
    u_s^+(t) = \frac{u_s(t) + u_s^\ast(t)}{\sqrt 2} = \frac{\cos(s \log(t))}{\sqrt{\pi t}} \,, \\
    u_s^-(t) = \frac{u_s(t) - u_s^\ast(t)}{\sqrt 2 i} = \frac{\sin(s \log(t))}{\sqrt{\pi t}} \,,
\end{align}
with $s \in \mathbb R^+$.
In this basis Carleman's operator becomes
\begin{equation}
\begin{split}
    \H(\omega,\omega') =
    \int_{s \in \mathbb R^+} \, \vert \lambda_s \vert^2 \big[ & u_s^+(\omega) u_s^+(\omega') \\ & + u_s^-(\omega) u_s^-(\omega')\big]  \,,
\end{split}
\end{equation}
making more explicit the (known) double degeneracy of its spectrum.

\def\gt{\widetilde g}
\def\rt{\widetilde r}
\vskip 2ex
\textbf{\textit{Least-square solution}} --- As mentioned in the main text, the coefficients $g_\alpha(t|\omega)$ are equivalently defined from the minimization of a least-square problem. 
Let us introduce new unknown coefficients $\gt_\alpha(t\vert\omega)$ and the (regularized) norm 
\begin{equation}
    \int_{\omega'} \left[ \delta(\omega-\omega')
    - \int_t e^{-\omega' t} \gt_\alpha(t|\omega) \right]^2
    + \alpha \int_t \gt_\alpha(t\vert\omega)^2 \,.
    \label{eq:least_square}
\end{equation}
By setting the first derivative w.r.t. $\gt_\alpha(t|\omega)$ to zero one finds
\begin{equation}
    \gt_\alpha(t|\omega) = \int_{t'} \, A_\alpha^{-1}(t,t') \, e^{-\omega t'}
    \,,
    \label{eq:gt_BG}
\end{equation}
where $A_\alpha^{-1}(t,t')$ is the inverse of the regularized operator
\begin{equation}
    A_\alpha(t,t') = \frac{1}{t+t'} + \alpha \, \delta(t-t') = \langle t \vert S_\alpha \vert t' \rangle
    \label{eq:A}
\end{equation}
which proves the equivalence with our definition in \eq{eq:gt}, namely $g_\alpha(t|\omega) = \gt_\alpha(t|\omega)$.

Similarly $\rho_\alpha(\omega)$ minimizes the distance with $\rho(\omega)$. In analogy to \eq{eq:least_square}, starting from a new set of unknown coefficients $\rt_\alpha(t)$, with little additional algebra one may prove that at the minimum of the following norm
\begin{equation}
    \int_\omega \Big[ \rho(\omega) - \int_t e^{-\omega t} \rt_\alpha(t) \Big]^2 + \alpha \int_t  \rt_\alpha(t)^2 \,,
\end{equation}
 $\rho_\alpha(\omega) = \int_t e^{-\omega t} \rt_\alpha(t)= \int_t g_\alpha(t|\omega) C(t)$.

\vskip 2ex
\textbf{\textit{Finite temporal extent}} --- At fixed $\alpha>0$, the systematic error due to the restriction of the time integral to $t\in[0,\tmax]$ of the continuous solution in \eq{eq:rhosol} is
\begin{equation}
    \rho_\alpha(\omega) - \rho_{\alpha,\tmax}(\omega) = \int_s \frac{u_s^\ast(\omega) \lambda_s}{\vert \lambda_s \vert^2 + \alpha} \int_{\tmax}^\infty \!\!\!\! dt \, u_s^\ast(t) C(t) \,.
    \label{eq:syst_tmax}
\end{equation}
From $|C(t)| \leq |C(\tmax)| e^{-M(t-\tmax)}$ for $t \geq \tmax$, we find
\begin{equation}
    \left\vert \int_{\tmax}^\infty dt \, C(t) u_s^\ast(t) \right\vert \leq \frac{|C(\tmax)|}{\sqrt{2 \pi M} \sqrt{M \tmax}} \,,
\end{equation}
which leads, with little additional algebra, to
\begin{equation}
\begin{split}
    \vert \rho_\alpha(\omega) - \rho_{\alpha,\tmax}(\omega) \vert
    \leq & \, \frac{|C(\tmax)|}{2 \pi \sqrt{\omega M} \sqrt{M \tmax}} \\ \times &  \int_s \frac{\vert\lambda_s\vert}{\vert\lambda_s\vert^2+\alpha} \,.
\end{split}
\end{equation}
This bound is finite as long as $\alpha>0$ and exponentially suppressed by $|C(\tmax)|$. Instead, from the concrete assumption $C(t) = c \, e^{-Mt}$ at $t\geq \tmax$, we obtain
\begin{equation}
    \int_{\tmax}^\infty dt \, u_s^\ast(t) C(t)
    = c \, u_s(M) \lambda_s^\ast(M \tmax) \,,
\end{equation}
where $\lambda_s^\ast(x) \equiv \Gamma\left(\frac12-is,x \right)$ is the upper incomplete $\Gamma$-function.
The resulting systematic error 
\begin{equation}
\begin{split}
    \rho_\alpha(\omega) - \rho_{\alpha,\tmax}(\omega) = c \int_s \frac{u_s^\ast(\omega) \lambda_s u_s(M)}{\vert\lambda_s\vert^2 + \alpha} \lambda_s^\ast(x) \,,
\end{split}
\end{equation}
is bounded and exponentially suppressed in $x=M\tmax$, and it can be calculated explicitly provided that $M$ and $c$ are known.

\vskip 2ex
\textbf{\textit{Properties of the $v_s(\omega,a)$ functions}} --- In this section we examine known properties of the so-called Hilbert matrix to derive the explicit functional form of the $v_s(\omega,a)$ functions in \eq{eq:vs}. 
We begin from the diagonalization of the (infinite) Hilbert matrix~\cite{Hill} regarded as an operator acting on $\ell^2(\mathbb Z^+)$
\begin{equation}
    \sum_{n=0}^\infty \frac{1}{n+m+2} x^\mu_n = \frac{\pi}{\sin(\pi \mu)} x^\mu_m
    \label{eq:hilbert}
\end{equation}
with $\mu=\frac12+is$, by the eigenfunctions 
\begin{equation}
    x^\mu_n = \sum_{k=0}^n \binom{n}{k} (-1)^k \frac{\Gamma(k+\mu) \Gamma(k+1-\mu)}{\Gamma(k+2)\Gamma(k+1)} \,.
    \label{eq:xn}
\end{equation}
By using a known integral representation of Euler's $\beta$ function and its relation with the $\Gamma$ function
\begin{equation}
    \beta(x,y) = \int_0^1 dr \, r^{x-1} (1-r)^{y-1} = \frac{\Gamma(x) \Gamma(y)}{\Gamma(x+y)} \,,
\end{equation}
and the property
\begin{equation}
    \sum_{k=0}^n \binom{n}{k} x^k = (1+x)^n \,, 
\end{equation}
with little algebra we derive the following representation in terms of the hypergeometric function ${}_3F_2$~\cite{Gradshteyn:1702455}
\begin{equation}
    x^\mu_n = \frac\pi{\sin(\pi\mu)} \ {}_3F_2 \left( \begin{array}{c|}
      -n, \mu, 1-\mu \\
       1,2 
    \end{array} \, 1 \right) \,,
\end{equation}
much more suitable for numerical evaluation than \eq{eq:xn}.
For $\mu=\frac12 + is$ the eigenvalues in \eq{eq:hilbert} coincide with those in the continuum $\vert \lambda_s \vert^2$. Instead the eigenfunctions $x_n^{\frac12+is}$ are related to the so-called continuous dual Hahn polynomials, defined according to~\cite{hahn1,hahn2,hahn3}
\begin{equation}
\begin{split}
    S_n(x^2;a,b,c) =  & \, (a+b)^{(n)} (a+c)^{(n)} \\ & \times {}_3F_2 \left( \begin{array}{c|}
      -n,a+ix,a-ix  \\
       a+b,a+c 
    \end{array} \, 1 \right) \,,
    \label{eq:hahn_poly}
\end{split}
\end{equation}
with the rising factorial being 
\begin{equation}
    (x)^{(n)} = \frac{\Gamma(x+n)}{\Gamma(x)} \,, \quad \forall x \in \mathbb C / \mathbb Z_0^-\,.
\end{equation}
They satisfy the orthogonality relation (see e.g. Ref.~\cite{hahn1})
\begin{equation}
\begin{gathered}
    \frac1{2\pi} \int_0^\infty ds \, \vert W_s(a,b,c) \vert^2 \, S_m(s^2;a,b,c) \, S_n(s^2;a,b,c) \\
    = \Gamma(n+a+b) \Gamma(n+a+c) \Gamma(n+b+c) n! \, \delta_{nm} \,,
    \label{eq:ortho1}
\end{gathered}
\end{equation}
with
\begin{equation}
    W_s(a,b,c) = \frac{\Gamma(a+is) \Gamma(b+is) \Gamma(c+is)}{\Gamma(2is)} \,.
\end{equation}
By setting $a=b=1/2$ and $c=3/2$, we obtain a completeness relation for the eigenfunctions of the Hilbert matrix acting on $\ell^2(\mathbb Z^+)$
\begin{equation}
    \int_0^\infty ds \, \frac{\vert \lambda_s \vert^2}{\vert N_s \vert^2} x_n^{\frac12+is} \, x_m^{\frac12+is} = \delta_{nm} 
    \label{eq:xmu_comp}
\end{equation}
with $N_s$ defined in \eq{eq:Ns}. 
Since the Hilbert matrix is typically expressed in terms of integer numbers, we denote its corresponding dimensionful counterpart as
$\overline A_a$, with matrix elements
\begin{equation}
    \overline A_a(an,am) = \frac{1}{a(n+m+2)} \,.
\end{equation}
It may be interpreted as the projection of \eq{eq:A} (at $\alpha=0$) to discrete time coordinates and its normalized eigenfunctions are given by
\begin{equation}
    \vbar_s(an,a) \equiv  \frac{\vert \lambda_s \vert}{\sqrt a \vert N_s \vert} x_{n}^{\frac12 + is} \,.
\end{equation}
Starting from the relation between $\overline A_a(t,t')$ and $\Hbar_a(\omega,\omega')$
\begin{equation}
\begin{split}
    a \sum_{n=0} e^{-a \omega(n+1)} & \overline A_a(an, a m) = \\ & \, 
    \int_{\omega'} \, \Hbar_a(\omega, \omega') e^{-a \omega' (m+1)} \,,
\end{split}
\end{equation}
one easily finds that they share the same spectrum 
\begin{equation}
    \int_{\omega'} \Hbar_a(\omega,\omega')  v_s(\omega',a) = |\lambda_s|^2 \,  v_s(\omega,a) \,,
\end{equation}
while their eigenfunctions are related as follows
\begin{equation}
    \vert \lambda_s \vert \, v_s(\omega,a) = a \sum_{n=0}^\infty e^{- a \omega (n+1)} \vbar_s(an,a) \,.
    \label{eq:vs_vsbar}
\end{equation}
From the exclusion of the first time slice $t=0$, it follows that $an = t-a$ as in \eq{eq:gtalpha_bar}.
Their orthogonality stems from the spectral theorem, while their completeness follows from \eq{eq:xmu_comp}.
Hence, they define a complete orthonormal basis in $\Lrho$ where $\Hbar_a(\omega,\omega')$ is diagonal
\begin{equation}
    \Hbar_a(\omega,\omega') = \int_0^\infty ds \, v_s(\omega,a) \vert \lambda_s \vert^2 v_s(\omega',a)\,.
\end{equation}
Now, we briefly study the leading discretization effects of the $v_s(\omega,a)$ functions, while a complete derivation will be reported in a longer publication~\cite{inprep}.
By starting from \eq{eq:vs_vsbar}, after some algebra, we arrive at the integral representation ($z = e^{-a\omega}[1- e^{-a\omega}]^{-1}$)
\begin{equation}
    v_s(\omega,a) = \sqrt a z \frac{ |\lambda_s|^2 }{|N_s|}
    {}_2F_1 \left( \begin{array}{c|}
      \frac12+is,\frac12-is    \\
       2
    \end{array} \, -z \right) \,.
    \label{eq:ymu_v1}
\end{equation}
Using known relations among the hypergeometric function ${}_2F_1$ with different arguments~\cite{Gradshteyn:1702455},
we obtain the equivalent form used in \eq{eq:vs}.
From the expansion of \eq{eq:vs} around $a=0$, we prove that discretization errors linear in the lattice spacing vanish, and in the continuum limit the functions 
$v_s(\omega,a)$ tend to the specific linear combination of $u_s(\omega)$ and $u_s^\ast(\omega)$ given in \eq{eq:cont_lim_vs}.
In the alternative basis of real functions in the continuum, $v_s(\omega,a)$ would tend to a combination of $u_s^+(\omega)$ and $u^-_s(\omega)$.
Finally, we note that by squaring \eq{eq:cont_lim_vs} we obtain four terms: $a[u_s(a\omega) u_s^\ast(a\omega') + u_s^\ast(a\omega) u_s(a \omega')]$ lead to $\delta(\omega-\omega')$ upon integration over $s$, while the remaining additional terms, proportional to $u_s(a\omega) u_s(a\omega')$ and $u_s^\ast(a\omega) u_s^\ast(a \omega')$, produce a vanishing contribution for $a\to 0$, after integrating in $s$.

\bibliography{biblio,bibliofv}

\end{document}